\documentclass[aps,pra,twocolumn,superscriptaddress,notitlepage]{revtex4-2}

\usepackage{amsmath,amssymb,amsfonts,mathrsfs}
\usepackage{physics}
\usepackage{graphicx}
\usepackage{bm}
\usepackage{tikz}
\usepackage[colorlinks=true,citecolor=red,linkcolor=blue,urlcolor=blue]{hyperref}
\usepackage{orcidlink}
\usetikzlibrary{arrows.meta,positioning,calc}

\begin{document}

\title{Measuring Entanglement in Qubit System}

\author{Vignesh S\,\orcidlink{0009-0000-6928-8426}}
\email{vickyvignesh2772@gmail.com}
\affiliation{Department of Physics, Jamia Millia Islamia, New Delhi-110025, India}
\author{Tabish Qureshi\,\orcidlink{0000-0002-8452-1078}}
\email{tqureshi@jmi.ac.in}
\affiliation{Centre for Theoretical Physics, Jamia Millia Islamia, New Delhi-110025, India}

\begin{abstract}

An operational way of measuring entanglement in a balanced two-path interferometers is presented, where path information is carried by some internal degree of freedom which, in turn, gets entangled with an ancilla system. The analysis is based on a tripartite description involving paths, an internal qubit degree of freedom, and some ancillary states entangled with the internal degree of freedom.  It is then applied to two physically distinct experimental situations: a modified Stern-Gerlach interferometer with spin-1/2 particles and a Mach-Zehnder interferometer with photons carrying polarization.
The ancilla degree of freedom may not be experimentally accessible.
Tracing out the ancillary system, and employing a quantum erasing procedure based on the internal degree of freedom, it is demonstrated that a concurrence-based measure of the entanglement, between the internal degree of freedom and the ancilla, can be extracted directly from measurable asymmetry of the two output channels. These results show that loss of coherence, quantum erasure, and entanglement estimation in interferometric experiments arise from the same underlying correlation structure and provide a compact experimentally accessible framework for quantifying entanglement in qubit systems.

\end{abstract}

\maketitle

\section{Introduction}
\begingroup
\setlength{\parindent}{0pt}
\setlength{\parskip}{0.7em}

The dual wave and particle nature of quantum systems has long been a central theme in the development of quantum mechanics. This duality, formalized through Bohr’s principle of complementarity \cite{Starke2026}, asserts that different experimental arrangements reveal mutually exclusive aspects of a quantum system. Interference experiments, such as the two-slit setup, provide a natural framework to explore this principle, where the visibility of interference fringes is directly linked to the indistinguishability of alternative paths.

It is well understood that the acquisition of which-path information leads to the suppression of interference. Even in the absence of an actual measurement, the mere possibility of distinguishing the paths, through correlations with additional degrees of freedom, is sufficient to destroy interference. This phenomenon is now commonly understood in terms of entanglement between the system and auxiliary ``which-path'' markers, leading to loss of coherence in the observed subsystem.

The concept of a quantum eraser \cite{Scully1982} provides a deeper insight into this interplay. By suitably choosing the measurement basis of the auxiliary system, it is possible to erase the which-path information and recover interference in selected subensembles. This has led to a wide range of theoretical proposals and experimental realizations in different physical systems, highlighting the fundamental role of correlations in determining observable behavior \cite{Zeilinger2016,Qureshi2025}.

Entanglement, which underlies these phenomena, is a key resource in quantum mechanics and quantum information theory. Quantifying entanglement in a physically meaningful and experimentally accessible way remains an important problem. For bipartite two-level systems, measures such as concurrence \cite{concurrence} provide a convenient way to characterize entanglement. However, establishing a direct connection between such measures and observable quantities, e.g. in interferometric experiments, is not always straightforward. It is therefore desirable to identify operational schemes in which entanglement can be assessed directly from measurable interference-related quantities.
While the quantitative relationship between which-path information, entanglement, and interference is well established, it is of considerable interest to formulate this connection in an experimentally accessible manner. In particular, extracting an entanglement measure from real experiments, remains an important problem.

In this work, we address this issue by analyzing two distinct interferometric systems: a reversible Stern--Gerlach interferometer for spin-$\tfrac{1}{2}$ particles and a Mach--Zehnder interferometer for photons carrying polarization. Despite their different physical realizations, both systems admit a common description in which path-information is encoded through correlations with auxiliary marker states.
We show that the overlap of the marker states provides a unified parameter governing coherence, interference visibility, and entanglement. Using this framework, we consider concurrence and demonstrate that it can be expressed in an operational form directly in terms of measurable detector probabilities. The same relation emerges in both Stern--Gerlach and Mach--Zehnder configurations, establishing a unified perspective on the loss of coherence, quantum erasure, and entanglement in qubit interferometric systems.

\endgroup

\section{Stern--Gerlach Interferometer}
\begingroup
\setlength{\parindent}{0pt}
\setlength{\parskip}{0.7em}

The Stern--Gerlach implementation studied here is a spin-path interferometer for spin-$\tfrac{1}{2}$ particles. Its purpose is to use the spin to keep track of the path information, recombine the two alternatives, and test whether the lost coherence can be recovered through a complementary readout. In this sense, it serves as the matter-wave analogue of a two-path interferometer with an internal marker degree of freedom. The sequence of elements used in this construction is shown in Fig.~\ref{fig:sg_setup}.
This setup has been extensively discussed in the original problem posed by Englert, Schwinger and Scully \cite{Humpty1988}.

\begin{figure}[htbp]
\centerline{\resizebox{8.5cm}{!}{\includegraphics{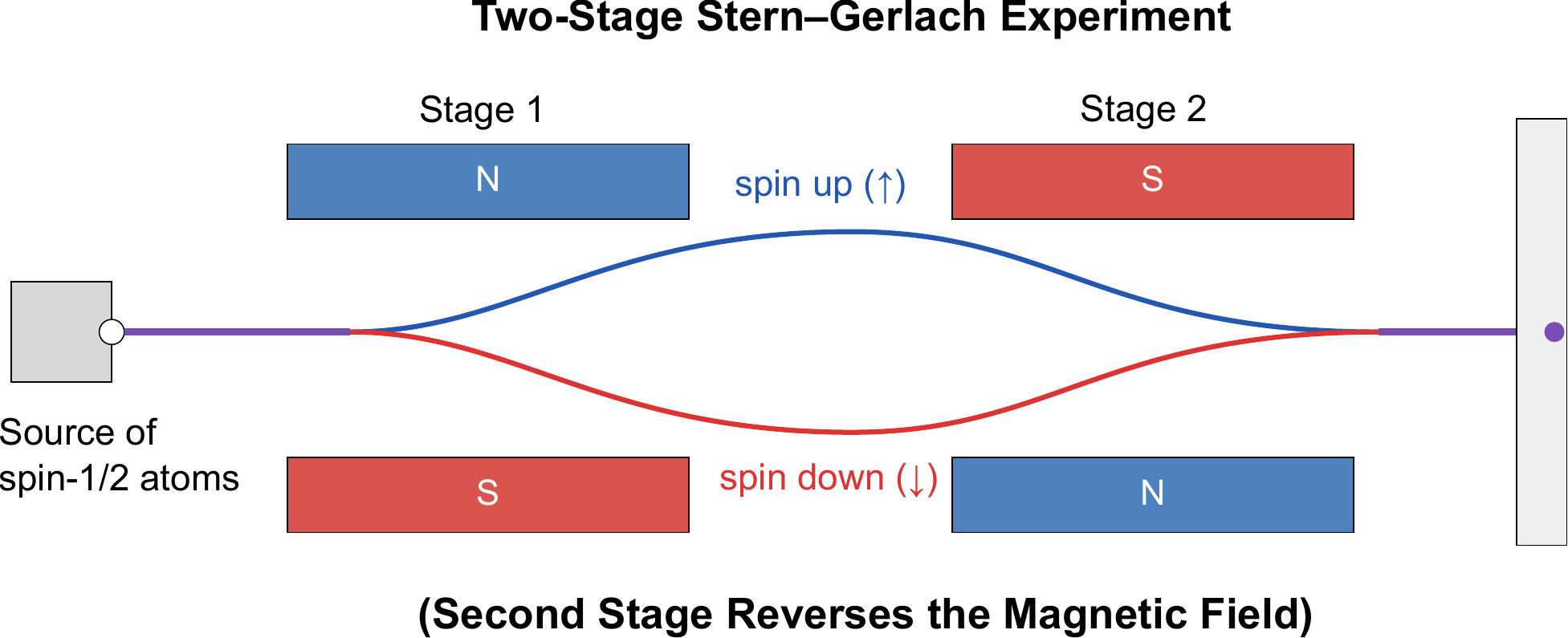}}}
\caption{Modified Stern-Gerlach interferometer for using the spin states as path markers. The spin, in addition, is entangled with an ancilla qubit. Another Stern-Gerlach setup is employed after this stage (no shown) with field in the x-direction, to implement quantum erasure before final detection.}
\label{fig:sg_setup}
\end{figure}

The state evolution begins with an incident beam prepared in an equal superposition of $\hat S_z$ eigenstates,
\begin{equation}
\ket{\Psi_0}=\ket{\psi_0}\otimes \frac{1}{\sqrt{2}}\left(\ket{+}+\ket{-}\right)\otimes \ket{m_0}.
\end{equation}
After the first $\hat S_z$ stage and the inhomogenous magnetic field pointing along the z-direction, the two separated paths become correlated with the two spin states. We assume that during its traversal the spin of the particle gets entangled to an ancillary system. The ancilla states $\ket{\alpha}$ and $\ket{\beta}$ get correlated to the two spin state, and the resultant state is given by
\begin{equation}
\ket{\Psi_1}=\frac{1}{\sqrt{2}}\left(\ket{\psi_{+}}\ket{+}\ket{\alpha}+\ket{\psi_{-}}\ket{-}\ket{\beta}\right).
\label{eq:sg_tagged}
\end{equation}
The states $\ket{\alpha},\ket{\beta}$ need not be orthogonal, though they are assumed to be normalized.
An ideal reversed Stern-Gerlach field recombines the two spatial branches into a common output mode $\ket{\psi_r}$, so that the combined state becomes
\begin{equation}
\ket{\Psi_2}=\ket{\psi_r}\otimes \frac{1}{\sqrt{2}}\left(\ket{+}\ket{\alpha}+\ket{-}\ket{\beta}\right).
\label{eq:sg_recombined}
\end{equation}
At this stage the spatial wave packets combine, and become disentangled with the spin and the ancilla. However the spin and the ancilla remain entangled.

The quantum eraser is implemented by measuring the spin in the complementary $\hat S_x$ basis, by letting the particle pass through a Stern-Gerlach field \emph{pointing along the x-direction}. A quantum eraser using a modified Stern-Gerlach setup has already been proposed \cite{QureshiRahman2012,Barney2019}.  The beam splits into two again, correlated to the states
\begin{equation}
\ket{\pm_x}=\frac{1}{\sqrt{2}}\left(\ket{+}\pm \ket{-}\right).
\end{equation}
In this basis, Eq.~\eqref{eq:sg_recombined} becomes
\begin{equation}
\ket{\Psi_3}=\ket{\psi_r}\otimes \frac{1}{2}\left[\ket{+_x}\left(\ket{\alpha}+\ket{\beta}\right)+\ket{-_x}\left(\ket{\alpha}-\ket{\beta}\right)\right].
\label{eq:sg_recombinedx}
\end{equation}
After passing through the eraser magnetic field, the spatial degrees of freedom of the particle again become correlated with the spin state, and the state has the following form
\begin{equation}
\ket{\Psi_4}= \tfrac{1}{2}\left[\ket{+_x}\left(\ket{\alpha}+\ket{\beta}\right)\ket{\psi_{x+}}+\ket{-_x}\left(\ket{\alpha}-\ket{\beta}\right)\ket{\psi_{x-}}\right].
\label{eq:sg_eraser}
\end{equation}
The states $\ket{\psi_{x+}},\ket{\psi_{x-}}$ represent two spatially separated beams, where the time-dependence is not explicitly depicted, being irrelevant to the analysis.
The states $\ket{\psi_{x+}},\ket{\psi_{x-}}$ will register at different spatial locations with probabilities $P_{1x}, P_{2x}$, respectively. These probabilities are given by
\begin{equation}
P_{1x}=\frac{1}{2}\left(1+\Re\langle\alpha|\beta\rangle\right),
\qquad
P_{2x}=\frac{1}{2}\left(1-\Re\langle\alpha|\beta\rangle\right),
\label{eq:sg_probs}
\end{equation}
which imply the channel asymmetry
\begin{equation}
\frac{P_{1x}-P_{2x}}{P_{1x}+P_{2x}}=\Re\langle\alpha|\beta\rangle.
\label{eq:sg_asymx}
\end{equation}
Thus the overlap of the ancilla states controls the recovered contrast after the eraser process. In this modified Stern--Gerlach setup, the experimentally accessible result is related to the overlap of the aniclla states, by the relation in Eq.~\eqref{eq:sg_asymx}. 

Furthermore, if one were to implement quantum eraser by letting the particle pass through a Stern-Gerlach field \emph{pointing along the y-direction} instead of x-direction, the relevant states with which the beams get correlated are given by
\begin{equation}
\ket{\pm_y}=\frac{1}{\sqrt{2}}\left(\ket{+}\pm i\ket{-}\right).
\end{equation}
A straightforward analysis, along the lines of the preceding one, leads one to the following channel asymmetry
\begin{equation}
\frac{P_{1y}-P_{2y}}{P_{1y}+P_{2y}}=\Im\langle\alpha|\beta\rangle.
\label{eq:sg_asymy}
\end{equation}

The connection of these results to entanglement is deferred to the unified treatment in Sec.~\ref{sec:unified}.

\endgroup

\section{Mach--Zehnder Interferometer}
\begingroup
\setlength{\parindent}{0pt}
\setlength{\parskip}{0.7em}

In the following we will show that a similar procedure can be set up in an optical setting, as a qubit is more often implemented by the polarization of a photon.  
In the optical setting, the relevant degrees of freedom are path, polarization, and ancilla states. A polarizing beam splitter first splits the incoming photon into two spatially distinct paths, correlating them with orthogonal polarization states (see Fig. \ref{fig:mzi-schematic}). A second balanced beam splitter then recombines the two paths, and splits them again, sending them to two different detectors. No interference will be seen in such a situation because the polarization states carry the path information of the photon, and complementarity forbids any interference. 
However, before detection, the path information can be erased by letting the photon pass through polarizers which do not discriminate between the two polarization states that the paths are entangled with. Such a procedure recovers interference, and the process is called quantum erasure. 

On the other hand, during its traversal of the two paths, the polarization of the photon may get entangled with an ancilla qubit. Or the polarization may already be entangled with the ancilla when the photon enters the interferometer. That is the situation we are interested in.
\begin{figure}[htbp]
\centerline{\resizebox{8.5cm}{!}{\includegraphics{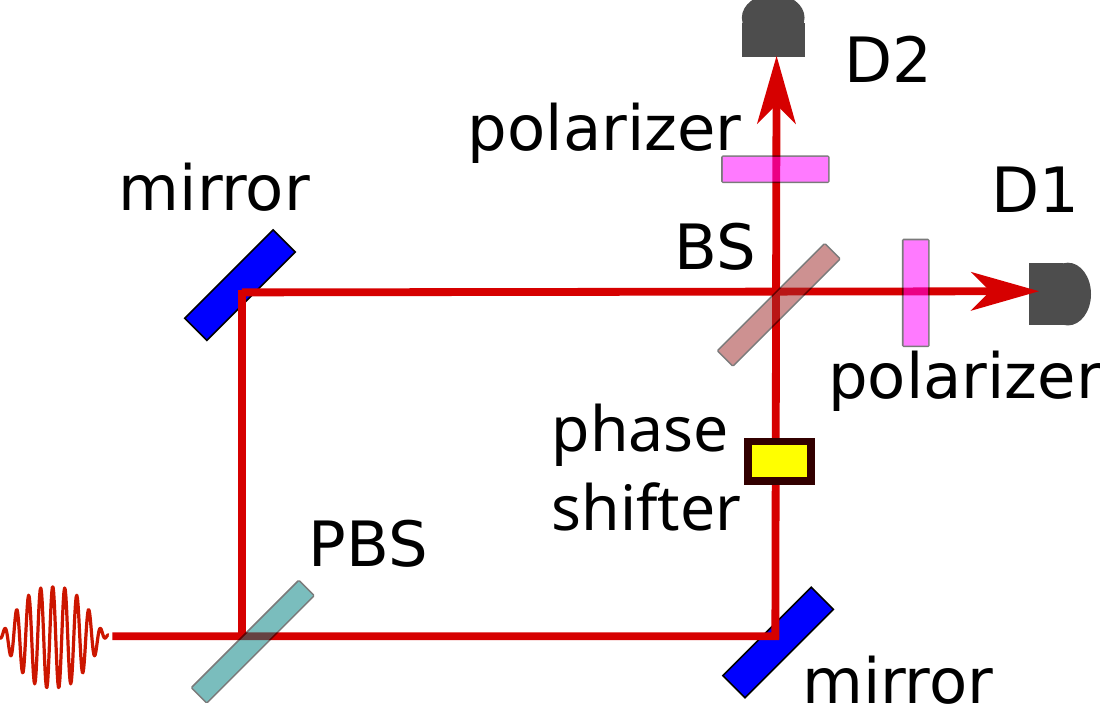}}}
\caption{Mach--Zehnder interferometer geometry. PBS1 creates two path amplitudes $|\psi_1\rangle,|\psi_2\rangle$, BS2 recombines them, and output polarizers enable quantum-eraser post-selection before detection at $D_1,D_2$.}
\label{fig:mzi-schematic}
\end{figure}
Starting from a single input mode $\ket{\psi_0}$, the initial polarization state is taken to be an equal superposition of the horizontal and vertical states. The combined initial state of the photon, the polarization, and the ancilla is given by
\begin{equation}
\ket{\Psi_0}=\ket{\psi_0}\otimes \frac{1}{\sqrt{2}}\left(\ket{H}\ket{\alpha}+\ket{V}\ket{\beta}\right).
\label{Psi0}
\end{equation}
After the first polarizing beam splitter, the horizontal and vertical components are routed into distinct paths, giving the tagged tripartite state
\begin{equation}
\ket{\Psi_1}=\frac{1}{\sqrt{2}}\left(\ket{\psi_1}\ket{H}\ket{\alpha}+\ket{\psi_2}\ket{V}\ket{\beta}\right).
\label{eq:mzi_tagged}
\end{equation}
Even when the entanglement happens after the photon enters the interferometer, the state is still described by the above relation.
Propagation through the second beam splitter is described by the standard balanced transformation
\begin{eqnarray}
U_{\mathrm{BS2}}\ket{\psi_1}&=&\tfrac{1}{\sqrt{2}}\left(\ket{D_1}+i\ket{D_2}\right) \nonumber\\
U_{\mathrm{BS2}}\ket{\psi_2}&=&\tfrac{1}{\sqrt{2}}\left(i\ket{D_1}+\ket{D_2}\right),
\label{eq:mzi_bs}
\end{eqnarray}
where $\ket{D_1},\ket{D_2}$ are the spatial photon states reaching detectors $D_1, D_2$, respectively. The output state then assumes the form
\begin{eqnarray}
\ket{\Psi_2} &=& U_{\mathrm{BS2}}\ket{\Psi_1}\nonumber\\
&=&\tfrac{1}{2}\Bigl[\ket{D_1}\bigl(\ket{H}\ket{\alpha}+i\ket{V}\ket{\beta}\bigr)\nonumber\\
&&+\ket{D_2}\bigl(i\ket{H}\ket{\alpha}+\ket{V}\ket{\beta}\bigr)\Bigr].
\label{eq:mzi_output}
\end{eqnarray}
Before any polarization postselection, the probabilities of the photon landing at the two detectors are equal,
\begin{equation}
P_{D_1}=P_{D_2}=\frac{1}{2}.
\label{eq:mzi_unconditional}
\end{equation}
Thus, as in the Stern--Gerlach case, recombination alone does not restore visible interference in unconditional counts because the which-path information is still encoded in correlated non-path degrees of freedom.

The quantum eraser is implemented by filtering the photons by their polarization in the circular basis,
\begin{equation}
\ket{R}=\frac{1}{\sqrt{2}}\left(\ket{H}+i\ket{V}\right),
\qquad
\ket{L}=\frac{1}{\sqrt{2}}\left(\ket{H}-i\ket{V}\right).
\label{eq:mzi_circular}
\end{equation}
Projecting Eq.~\eqref{eq:mzi_output} onto the detector and polarization channels gives
\begin{align}
P_{D_1}^{(R)}&=\frac{1}{4}\left(1+\Re\langle\alpha|\beta\rangle\right), &
P_{D_1}^{(L)}&=\frac{1}{4}\left(1-\Re\langle\alpha|\beta\rangle\right), \label{eq:mzi_d1}\\
P_{D_2}^{(R)}&=\frac{1}{4}\left(1-\Re\langle\alpha|\beta\rangle\right), &
P_{D_2}^{(L)}&=\frac{1}{4}\left(1+\Re\langle\alpha|\beta\rangle\right). \label{eq:mzi_d2}
\end{align}
These conditional channels recover the overlap information through the asymmetries
\begin{equation}
\frac{P_{D_1}^{(R)}-P_{D_1}^{(L)}}{P_{D_1}^{(R)}+P_{D_1}^{(L)}}=\Re\langle\alpha|\beta\rangle,
\qquad
\frac{P_{D_2}^{(L)}-P_{D_2}^{(R)}}{P_{D_2}^{(L)}+P_{D_2}^{(R)}}=\Re\langle\alpha|\beta\rangle.
\label{eq:mzi_asymR}
\end{equation}
The optical quantum eraser therefore functions in exactly the same way as the Stern--Gerlach eraser. The asymmetry in the intensity in the two channels is related to the overlap of the two ancilla states. 

Instead of using the circular polarization basis, one may choose to use the diagonal-antidiagonal polarization basis
\begin{equation}
\ket{d_{\pm}}=\frac{1}{\sqrt{2}}\left(\ket{H}\pm\ket{V}\right).
\label{eq:mzi_diagonal}
\end{equation}
Carrying out an analysis on the preceding line, one finds that the asymmetries in the two channels are related to overlap of the two states as follows
\begin{equation}
\frac{P_{D_1}^{(d_+)}-P_{D_1}^{(d_-)}}{P_{D_1}^{(d_+)}+P_{D_1}^{(d_-)}}=\Im\langle\alpha|\beta\rangle,
\quad
\frac{P_{D_2}^{(d_-)}-P_{D_2}^{(d_+)}}{P_{D_2}^{(d_-)}+P_{D_2}^{(d_+)}}=\Im\langle\alpha|\beta\rangle,
\label{eq:mzi_asymD}
\end{equation}
where $P_{D_j}^{(d_{\pm})} = 
|\langle d_{\pm}|\langle D_j|~|\Psi_2\rangle|^2$

An objection may be raised here regarding the generality of the equal superposition form of the state (\ref{Psi0}). We are assuming an ensemble of photons with the polarization entangled with an ancilla. The polarization of the photon can always be rotated until an equal superposition of the entangled state is obtained, without affecting the entanglement of the state. This is possible if an ensemble of such particles is available.

As in the Stern--Gerlach case, the experimentally relevant result at this stage is the asymmetry relations in Eqs.~\eqref{eq:mzi_asymR} and \eqref{eq:mzi_asymD}. In the following section we discuss how a measure of entanglement can be extracted from these experiments.

\endgroup

\section{Unified approach to measuring entanglement}\label{sec:unified}
\begingroup
\setlength{\parindent}{0pt}
\setlength{\parskip}{0.7em}

The modified Stern--Gerlach and the modified Mach--Zehnder analyses can be written in a single common language. Although the two systems are physically different, they share the same information-theoretic structure: a coherent input is split into two alternatives, each alternative becomes correlated with an path marker qubit and an ancilla state, the spatial paths are recombined, and a complementary readout reveals the coherence stored in the correlations. 

The asymmetry relations obtained in Eqs.~\eqref{eq:sg_asymx} and \eqref{eq:mzi_asymR} can now be translated into a single entanglement measure. Instead of deriving separate concurrence formulas for each apparatus, it is more transparent to introduce the measure once in the common qubit--marker language and then read the Stern--Gerlach and Mach--Zehnder interferometers as two concrete realizations of the same result.

This common structure is captured by the effective two-party state
\begin{equation}
\ket{\Psi}=\frac{1}{\sqrt{2}}\left(\ket{0}\ket{\alpha}+\ket{1}\ket{\beta}\right),
\qquad
s\equiv \langle\alpha|\beta\rangle,
\label{eq:unified_state}
\end{equation}
where $\ket{0}$ and $\ket{1}$ denote the relevant two-level basis states and $s$ is the marker overlap. In the Stern--Gerlach realization, $\ket{0},\ket{1}$ correspond to $\ket{+},\ket{-}$ spin states; in the Mach--Zehnder realization, they correspond to $\ket{H},\ket{V}$ polarization states. Tracing over the ancilla subsystem gives the reduced qubit density matrix
\begin{equation}
\rho_{\mathrm{sys}}=\tfrac{1}{2}
\begin{pmatrix}
1 & s^{\ast}\\
s & 1
\end{pmatrix}.
\label{eq:unified_rho}
\end{equation}
Its off-diagonal terms contain the entire coherence information, so the magnitude $|s|$ directly determines how much local coherence survives in the observed subsystem.

For a pure bipartite two-level state, the concurrence is obtained from the purity of the reduced state. Using Eq.~\eqref{eq:unified_rho}, one finds
\begin{equation}
\Tr\!\left(\rho_{\mathrm{sys}}^2\right)=\frac{1}{2}\left(1+|s|^2\right),
\qquad
\mathscr{C}^2=1-|s|^2=1-\abs{\langle\alpha|\beta\rangle}^2.
\label{eq:unified_conc}
\end{equation}
Thus the same overlap parameter that governs reduced-state coherence also fixes the entanglement content. In the balanced pure-state limit, this gives the compact complementarity relation $\mathscr{C}^2+V^2=1$, with visibility $V=|s|$. With the two paths equally probable, this is already expected \cite{triality}.

To connect this result directly with experiment, consider the eraser basis
\begin{equation}
\ket{x_{\pm}}=\frac{1}{\sqrt{2}}\left(\ket{0}\pm \ket{1}\right).
\label{eq:eraser_xbasis}
\end{equation}
Projecting Eq.~\eqref{eq:unified_state} onto these channels yields
\begin{equation}
P_{x\pm}=\frac{1}{2}\left(1\pm \Re s\right).
\label{eq:unified_probs}
\end{equation}
The corresponding normalized asymmetry is therefore
\begin{equation}
A_x\equiv \frac{P_{x+}-P_{x-}}{P_{x+}+P_{x-}}=\Re s.
\label{eq:unified_xasym}
\end{equation}
Alternately, if one uses the eraser basis
\begin{equation}
\ket{y_{\pm}}=\frac{1}{\sqrt{2}}\left(\ket{0}\pm i\ket{1}\right),
\label{eq:eraser_ybasis}
\end{equation}
the corresponding normalized asymmetry turns out to be
\begin{equation}
A_y\equiv \frac{P_{y+}-P_{y-}}{P_{y+}+P_{y-}}=\Im s.
\label{eq:unified_yasym}
\end{equation}
These results can be applied to Eq.~\eqref{eq:unified_conc} yielding the operational entanglement measure
\begin{eqnarray}
\mathscr{C}^2 &=& 1-\abs{\langle\alpha|\beta\rangle}^2
= 1 - (\Re\langle\alpha|\beta\rangle)^2 - (\Im\langle\alpha|\beta\rangle)^2\nonumber\\
 &=& 1 - A_x^2 - A_y^2 ~.
\label{eq:unified_operational}
\end{eqnarray}
Consequently, concurrence may be quantified through the execution of the quantum eraser experiment utilizing two distinct bases, followed by an assessment of the output beam asymmetry across both scenarios. This methodological framework is precisely instantiated in both interferometric configurations: in the Stern-Gerlach apparatus via the dual spin-resolved eraser pathways, and in the Mach-Zehnder apparatus via the polarization-resolved detection channels. The various experimental implementations employ distinct mechanisms to isolate quantum states across different bases.

Three general conclusions follow from this unified description. First, decoherence is relational: the global state can remain pure while local coherence is lost through correlations. Second, quantum erasure is a basis-selection effect rather than a reversal of earlier dynamics. Third, entanglement estimation is operational: channel asymmetries measured in the appropriate basis can be converted directly into concurrence. These points show that the Stern--Gerlach and Mach--Zehnder systems are not merely analogous examples, but concrete realizations of the same general qubit--marker interferometric template.

\endgroup

\section{Conclusion}
\begingroup
\setlength{\parindent}{0pt}
\setlength{\parskip}{0.7em}

We examined loss of coherence, quantum erasure, and entanglement within a unified interferometric framework. The central strategy was to treat the path, the effective qubit degree of freedom, and the ancilla states as parts of a correlated quantum system, and then identify how measurable probabilities depend on the marker-state overlap. In both the Stern--Gerlach and Mach--Zehnder realizations, the suppression of visible interference was traced to correlation-induced loss of local coherence rather than to any nonunitary destruction of the global state. 

The analysis established a common quantitative structure across the two platforms. The overlap parameter $\langle\alpha|\beta\rangle$ determines the off-diagonal terms of the reduced state, governs the residual coherence of the observed subsystem, and fixes the entanglement content through the concurrence relation 
$\mathscr{C}^{2}=1-\abs{\langle\alpha|\beta\rangle}^{2}$.
For the setups considered here, the same quantity is obtained directly from the asymmetry of the eraser-resolved output probabilities, giving an operational route from measured detector counts to entanglement estimation. It may be pointed out that this result may be looked up as a consequence of the triality relation established earlier \cite{triality}, but doesn't actually use it. Rather it provides a general framework to extract concurrence in such experiments.

In the Stern--Gerlach case, spin-path tagging followed by $\hat S_x$ or $\hat S_y$ analysis produces two complementary postselected channels whose imbalance directly measures the marker overlap. In the Mach--Zehnder case, path-polarization tagging followed by circular-polarization analysis reproduces the same operational structure. These two systems are therefore physically distinct realizations of a single information-theoretic template in which which-path information, coherence loss, and entanglement are controlled by the same correlation parameter. One may think of alternate realizations of the concept discussed here, to experimentally measure concurrence. For example, neutron interference experiments have already been proposed in the Mach-Zehnder geometry \cite{neutrons2000}. 

The main conceptual result is that the loss of coherence, quantum erasure, and entanglement estimation are not separate phenomena, but different manifestations of a common underlying correlation structure. This provides a compact and experimentally meaningful framework for understanding qubit interferometers with path marking. Natural extensions of the present work include the treatment of unequal input amplitudes, mixed initial states, and explicit modeling of realistic experimental imperfections.

\endgroup

\begin{acknowledgments}
VS thanks the Department of Physics and the Centre for Theoretical Physics, Jamia Millia Islamia, for providing a supportive academic environment in which this study was carried out.
\end{acknowledgments}

\section*{Declarations}

\noindent No funding was received for conducting this study.
\vskip 3mm

\noindent The authors have no competing interests to declare that are relevant to the content of this article.
\vskip 3mm

\noindent No additional data was generated in this work.



\begin{thebibliography}{99}
\bibitem{Starke2026}
D. S. Starke, J. Maziero, M. L. W. Basso, and T. Qureshi,
"Bohr's complementarity,"
\href{https://doi.org/10.48550/arXiv.2605.26375}
{arXiv:2605.26375 [quant-ph]}

\bibitem{Scully1982} M.~O.~Scully and K.~Dr\"uhl,
``Quantum Eraser: A Proposed Photon Correlation Experiment Concerning Observation and `Delayed Choice' in Quantum Mechanics,''
\href{https://doi.org/10.1103/PhysRevA.25.2208}
{Phys. Rev. A {25}, 2208 (1982)}.

\bibitem{Zeilinger2016} X. Ma, J. Kofler, and A. Zeilinger, ``Delayed-choice gedanken experiments and their realizations'', 
\href{https://doi.org/10.1103/RevModPhys.88.015005}
{Rev. Mod. Phys. 88, 015005 (2016)}.

\bibitem{Qureshi2025}
T. Qureshi, ``The Enigma of Delayed Choice Quantum Eraser,'' 
\href{https://doi.org/10.12743/quanta.93}
{Quanta 14, 66 (2025)}. 

\bibitem{concurrence} S. Hill, W.K. Wootters, ``Entanglement of a pair of quantum bits,''
\href{https://doi.org/10.1103/PhysRevLett.78.5022}
{\emph{Phys. Rev. Lett.} \textbf{78}, 5022 (1997).}

\bibitem{Humpty1988} B.-G. Englert, J. Schwinger, and M. O. Scully, 
``Is spin coherence like Humpty-Dumpty? I. Simplified treatment,'' 
\href{https://doi.org/10.1007/BF01909939}
{Found. Phys. 18, 1045 (1988)}. 

\bibitem{QureshiRahman2012}
T.~Qureshi and Z.~Rahman,
\textit{Quantum eraser using a modified Stern-Gerlach setup},
\href{https://doi.org/10.1007/BF01909939}
{Prog. Theor. Phys. {127}, 71 (2012)}.

\bibitem{Barney2019}
R.~D.~Barney and J.-F.~S.~Van Huele,
\textit{Quantum coherence recovery through Stern-Gerlach erasure},
\href{https://doi.org/10.1088/1402-4896/ab2d45}
{Phys. Scr. \textbf{94}, 105105 (2019).}

\bibitem{triality} A. K. Roy, N. Pathania, N. K. Chandra, P. K. Panigrahi, and T. Qureshi, ``Coherence, path predictability, and I-concurrence: A triality,'' 
\href{http://dx.doi.org/10.1103/PhysRevA.105.032209}
{Phys. Rev. A 105, 032209 (2022).}

\bibitem{neutrons2000} G. Badurek, R. J. Buchelt, B.-G. Englert, and H. Rauch, 
``Wave–particle duality and quantum erasure in polarized–neutron interferometry,"
\href{https://doi.org/10.1016/S0168-9002(99)01037-2}
{Nuclear Instruments and Methods in Physics Research Section A 440, 562 (2000)}. 

\end{thebibliography}
\end{document}